\documentclass[%
reprint,
superscriptaddress,
nofootinbib,
amsmath,amssymb,
aps,
pra,
floatfix,
]{revtex4-1}

\usepackage{xfrac}
\usepackage{graphicx}
\usepackage{tensor}

\usepackage{braket}
\usepackage{color}
\usepackage{upgreek}

\definecolor{OliveGreen}{RGB}{0,200,200}

\begin{document}

\preprint{APS/123-QED}

\title{Entanglement generation via power-of-SWAP operations between dynamic electron-spin qubits}

\author{=Hugo V. Lepage}
\affiliation{
Cavendish Laboratory, Department of Physics, University of Cambridge, Cambridge CB3 0HE, United Kingdom
}%
\author{=Aleksander A. Lasek}%
\affiliation{
Cavendish Laboratory, Department of Physics, University of Cambridge, Cambridge CB3 0HE, United Kingdom
}%
\affiliation{Hitachi Cambridge Laboratory, J. J. Thomson Avenue, CB3 0HE, Cambridge, United Kingdom
}%
\author{David R. M. Arvidsson-Shukur}%
\affiliation{
Cavendish Laboratory, Department of Physics, University of Cambridge, Cambridge CB3 0HE, United Kingdom
}%
\affiliation{Department of Mechanical Engineering, Massachusetts Institute of Technology, Cambridge, Massachusetts 02139, USA
}%
\author{Crispin H. W. Barnes}
\affiliation{
Cavendish Laboratory, Department of Physics, University of Cambridge, Cambridge CB3 0HE, United Kingdom
}%

\date{\today}

\begin{abstract}

Surface acoustic waves (SAWs) can create moving quantum dots in piezoelectric materials. Here we show how electron-spin qubits located on dynamic quantum dots can be entangled. Previous theoretical and numerical models of quantum-dot entanglement generation have been insufficient to study quantum dynamics in realistic experimental devices. We utilize state-of-the-art graphics processing units to simulate the wave function dynamics of two electrons carried by a SAW through a 2D semiconductor heterostructure. We build a methodology to implement a power-of-SWAP gate via the Coulomb interaction. A benefit of the SAW architecture is that it provides a coherent way of transporting the qubits through an electrostatic potential. This architecture allows us to avoid problems associated with fast control pulses and guarantees operation consistency, providing an advantage over static qubits. For inter-dot barrier heights where the double occupation energy is sufficiently greater than the double-dot hopping energy, we find that parameters based on experiments in GaAs/AlGaAs heterostructures can produce a high-fidelity root-of-SWAP operation. Our results provide a methodology for a crucial component of dynamic-qubit quantum computing.


\end{abstract}

\maketitle


\section{Introduction}
The development of a universal semiconductor quantum computer hinges on the ability to entangle qubits. One promising method is to use the exchange interaction between electron-spins.
This concept was first introduced by Loss \emph{et al.}, for static qubits \citep{loss1998quantum,burkard1999coupled}, and Barnes \emph{et al.}, for dynamic, also called $\textit{flying}$,  qubits carried by surface acoustic waves (SAWs) \citep{barnes2000quantum}.
The use of flying qubits trapped in SAWs is a particularly  favorable platform for quantum computation for two reasons. First, the dynamic nature of the qubits enables on-chip operations to be controlled by static electric and magnetic fields from surface gates and magnetic microstructures \citep{Furuta2004,McNeil2010}. Not having to vary surface gate potentials reduces associated errors. Second, the confinement caused by the SAW potential prevents spatial dispersion of the fermionic wave packets  \citep{arvidsson2017protocol}. The framework is especially promising for building a universal quantum transducer---a bus that transports entangled qubits between spatially separated parts of a quantum computer that could itself be implemented in a different technology \citep{PhysRevX.5.031031, Salari2019}.

The last decade has seen significant developments in the achievement of SAW technologies \citep{divincenzo2000universal,malinowski2018spin}. Advances include the reliable control of single-electron transport \citep{mcneil2011demand, hermelin2011electrons, ford2017transporting, edlbauer2017non} and the increase in electron-qubit coherence times \citep{stotz2005coherent, sanada2013manipulation, lazic2014scalable, petta2005coherent}. The SAW framework for manipulating electron-spin qubits has shown promise for realizing optics-like quantum processes with readily interacting particles. Experimentally, single-qubit operations \citep{edlbauer2017non}, beam-splitters \citep{rodriquez2005surface,Takada2019} and spin-polarization readout devices \citep{kosaka2009spin} have been realized in GaAs heterostructures, and a spin-qubit toolkit for the implementation of generalized measurements has been presented \citep{arvidsson2017protocol}.

Previous works on flying electron systems have been restricted to either single particle scenarios \citep{Buscemi2010, Bordone2004}, or analytical models with a limited number of sites \citep{burkard1999coupled, barnes2000quantum} and simplified simulations in 1D for two particles\citep{owen2012generation, giavaras2007generation}.

In the latter case, it was suggested that entanglement generation could be achieved using a single-shot root-of-SWAP operation in which two electrons collide in a harmonic potential. See Appendix \ref{appendix:PowerOfSwap} for a definition of a logic power-of-SWAP operation.
Attempts simulate realistic devices in layered 3D systems have faced problems owing to the space- and time-domain scaling associated with solutions to the many-particle time-dependent Schr\"{o}dinger equation (TDSE). However, recent advances in graphics-processing-unit (GPU) performance \citep{Arunkumar2017,7979887} have made previously demanding problems readily solvable.

In this paper, we utilize state-of-the-art GPU hardware to run a customized $\textit{staggered-leapfrog}$ algorithm \citep{goldberg1967computer, askar1978explicit, maestri2000two}. 
Using a combination of two previous time steps to solve the next, alongside iterative updates of the real and imaginary parts of the wave function, enables us to simulate the dynamics of two interacting electrons efficiently and accurately.
In particular, we study SAW-based flying qubits, interacting via the Coulomb interaction in a 2D double-dot potential. The combination of our custom-built GPU hardware and tailored software allows us to simulate time-dependent quantum dynamics in a computation time on the order of days rather than years. Our results demonstrate the experimental viability of entanglement generation via root-of-SWAP operations. Furthermore, we show that the single-shot method \citep{owen2012generation} is not experimentally feasible. Not only are our simulations useful to gain insight into quantum logic operations, they also shed new light on simpler analytical models. Specifically, we compare our simulation results to two commonly used two-site models: the Hubbard approach \citep{barnes2000quantum}, and the Hund-Mulliken method for molecular orbitals \citep{burkard1999coupled}. We establish the limits and applicability of these models. We use experimentally realistic parameters for the interaction duration, the device potential, and geometry. 
To ensure parameter realism, we calculate the potential profile of the heterostructure with voltages applied to the metallic gates. We use a Poisson-Schr\"odinger self-consisted solver to calculate the range of values that are possible with current semiconductor technologies \cite{Takada2019}. Since this work demonstrates a proof-of-concept for the SAW-driven entangling operation, we use analytical equations to reproduce the potentials calculated by our solver. In doing so, we avoid simulating a specific device implementation and ensure that these simulations are reproducible and adaptable to different experimental needs.
Our work is a vital step towards constructing the fundamental building blocks of a SAW-based quantum computer.
The simulations we present are based on the GaAs/AlGaAs SAW-based heterostructure but our methodology, results, and conclusions are applicable to other semiconductor quantum systems, including static quantum dots.

This paper is structured as follows.
In Sec.\ \ref{sec:Device}, we describe in detail the semiconductor device we use as a model in our simulations of SAW-driven electron dynamics. 
In Sec.\ \ref{sec:Analytical}, we provide an analytical description of a device potential, as well as calculations of expected evolution of the two-particle wave function during the root-of-SWAP operation. 
In Sec.\ \ref{sec:SimResults}, we describe our numerical techniques and compare two methods for generating entanglement: the collision method and Coulomb tunneling method. We also compare these results to simple analytical two-site models.
Sec.\ \ref{sec:Sensitivity} contains a study of the sensitivity of the logic operation as a function of changes in the experimental parameters. 
Finally, in Sec.\ \ref{sec:Conclusion}, we conclude with a discussion of our results.

\section{Device Description}
\label{sec:Device}

Here we describe the device structure and the electron dynamics that allow us to model an entangling operation between two spin-qubits. The physical spin-qubits are electrons, and their spatial dynamics are controlled by SAWs. In each operation, two qubits travel through channels separated by a potential barrier. At the locus of the entangling operation, this barrier is lower, allowing the electron qubits to swap via the exchange interaction.

\begin{figure}[htpb]
\centering
\includegraphics[width=0.5\textwidth]{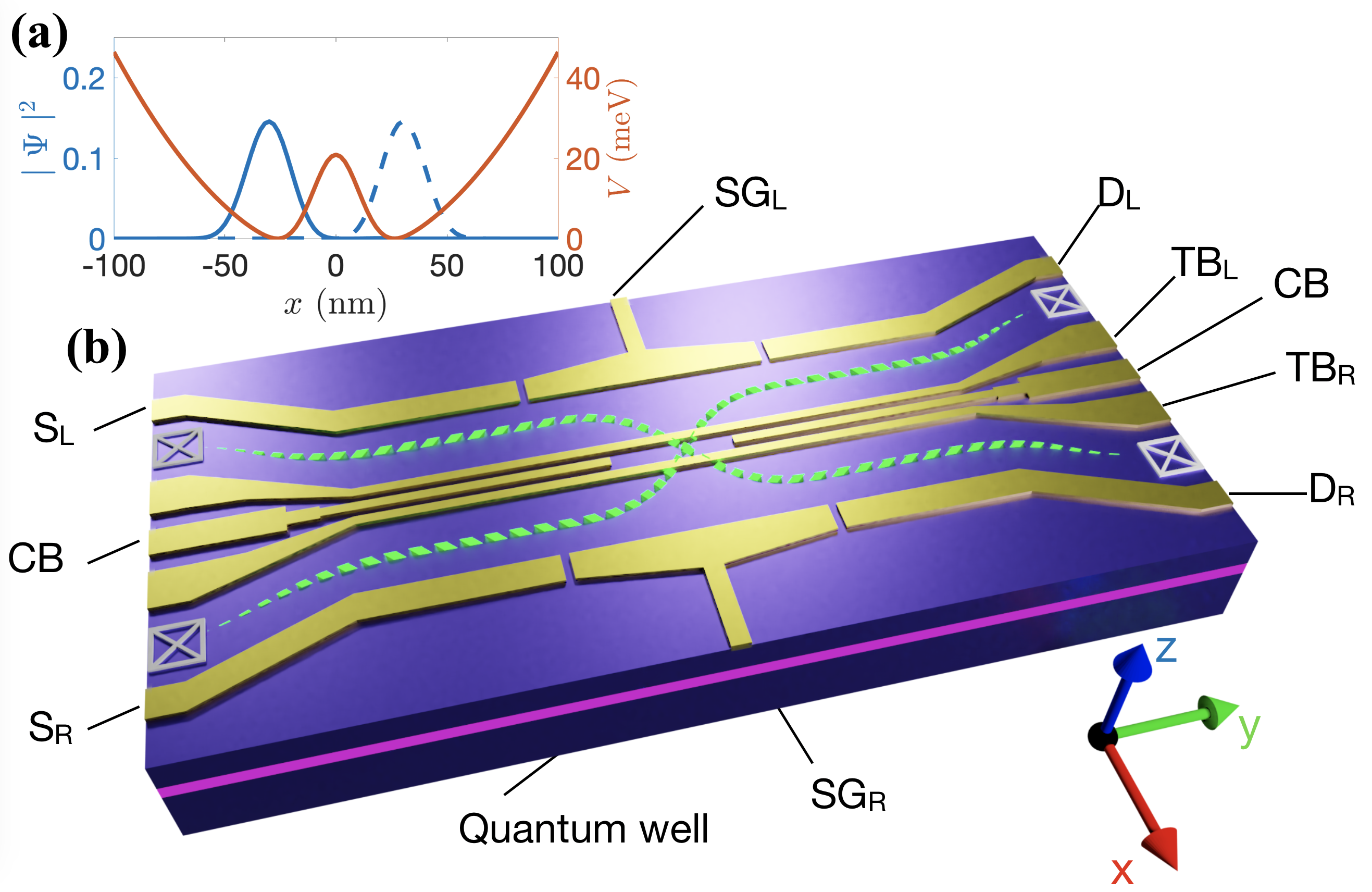}
\caption{ \textbf{(a)} Cross section of the potential layout in the region of high barrier alongside a trace of the initial state of the wave function along the $x$-dimension. \textbf{(b)} Schematic of a SAW-based power-of-SWAP device. Electrons are carried by SAWs from bottom to top (positive $y$-coordinate) along two channels, undergoing a power-of-SWAP operation in the central gate region. Green dotted lines show the path electrons can take through the device.}
\label{fig:Device}
\end{figure}

Fig.\ \ref{fig:Device} shows a SAW device designed to carry out a power-of-SWAP entangling operation on two electrons. The device is an adaptation of the one presented in \citep{barnes2000quantum}. Sinusoidal SAWs are generated by interdigitated transducers and propagate as transverse plane waves in the positive $y$-direction. The SAWs modulate the electric potential of a piezoelectric substrate to produce a train of quantum dots propagating along channels defined by metallic gates. The SAWs trap pairs of electrons from a two-dimensional electron gas in the same minimum \citep{Robinson2001}, with one electron in each channel (separated in the $x$-direction by the tunnel barrier). The SAW then carries the electrons through their respective channels. In the center of the device, where the barrier is lower, the tunneling rate can be controlled by voltages on $\text{TB}_{\text{L}}, \text{TB}_{\text{R}}$. 

The quantum dynamics of the system are generated by the travelling SAWs, therefore the voltages on the surface gates can be held constant throughout the entangling operation. This gives the SAW-based system a significant advantage over static qubit systems, which are controlled by generating voltage pulses that introduce charge noise and can induce stray SAWs, causing decoherence.

In what follows we will consider a device consisting of a GaAs/AlGaAs heterostructure containing a single layer of two-dimensional electron gas trapped in a quantum well.  On the top surface, a pattern of Schottky gates creates the two channels running in the $y$-direction, separated by a central barrier. We define the barrier's center as the origin of the $x$-direction and label the two channels with subscript $\text{L}$ (left) and $\text{R}$ (right) for negative or positive $x$, respectively. Negative voltages on the gates labelled by $\text{S}_{\text{L}}, ~ \text{S}_{\text{R}}, ~ \text{SG}_{\text{L}}, ~ \text{SG}_{\text{R}}, ~ \text{D}_{\text{L}}, ~ \text{D}_{\text{R}}$ generate the outer walls of the two channels.  Voltages on the gates labelled $\text{CB}$, $\text{TB}_{\text{L}}$, and $\text{TB}_{\text{R}}$ control the profile of the central barrier, and ensure that it strongly separates the two channels, except at the middle part of the device in the $y$-direction, where the barrier is lower. It is in the region of lower central barrier that the entangling operation occurs. $\text{TB}_{\text{L}}$ and $\text{TB}_{\text{R}}$ are sufficiently close that they only produce a single potential maximum, in the $x$-direction, between the two channels (Fig.\ \ref{fig:Device}(a)).


\section{Analytical Model}
\label{sec:Analytical}

To describe quantum dynamics in this device,  we use a two-particle Hamiltonian of the form
\begin{equation}
    \hat{H}=\sum_{i=1,2} \left(\frac{\hat{p}_i^2}{2m_i} + \hat{V}_{\text{D}}(r_i) +\hat{V}_{\text{SAW}}(t,r_i) \right) + \hat{V}_\text{C}(r_1,r_2),
    \label{eq:HamiltonianDevice}
\end{equation}
where $\hat{V}_\text{C}(r_1,r_2)$ is the two-particle Coulomb potential, $\hat{V}_{\text{SAW}}(t,r)$ is the SAW potential carrying the electrons along the channels and $\hat{V}_{\text{D}}(r)$ is the device potential.  This potential is made up of two parallel harmonic channels running along the $y$-dimension, coupled in the central gate region by a Gaussian tunnel barrier, forming a double quantum dot with harmonic confinement perpendicular to the channels, along the $x-$dimension. An explicit expression is given in Appendix \ref{appendix:Parameters}.
By boosting our reference frame to match the velocity of the SAW, which is constant, we can treat $\hat{V}_{\text{SAW}}(t,r)$ as a time-independent confining potential along the channel direction. 
Finding the eigenstates of the boosted time-independent Hamiltonian using a number basis derived from second quantization allows us to obtain the two-particle wave functions when the barrier between both channels is static. 
Since the potential does not have any explicit spin dependence, because of a weak Lorentz term, single-qubit spin rotations do not occur.

\begin{figure*}[htp]
\centering
\includegraphics[width=0.75\textwidth]{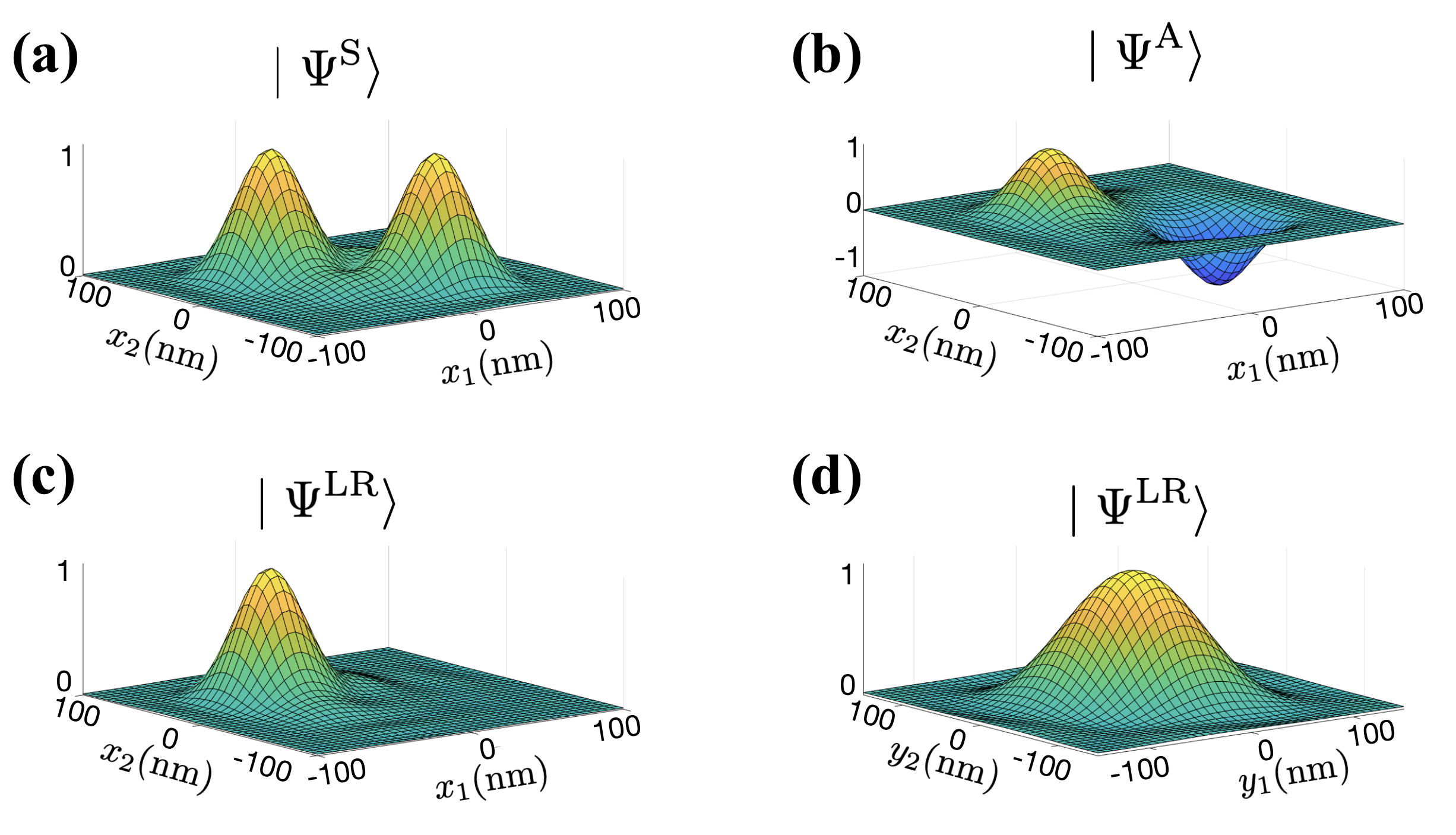}
\caption{\textbf{Two-particle spatial wave functions. }\textbf{ (a)} Ground state $\ket{\Psi^\text{S}}$ ($y_1 = 0$ and $y_2 = 0$). \textbf{(b)} First excited state $\ket{\Psi^\text{A}}$ ($y_1 = 0$ and $y_2 = 0$). \textbf{(c)} Combination of the ground state and first excited state $\ket{\Psi^{\text{LR}}}$. The first particle is localized in the left channel and the second particle is localized in the right channel ($y_1 = 0$ and $y_2 = 0$). \textbf{(d)} Gaussian spread of both particles in the $y$-dimension ($x_1 = 0$ and $x_2 = 0$). All four panels show the wave function divided by its extremum, with the z-axis in arbitrary units.}
\label{fig:InitialState}
\end{figure*}

We assume that the electrons in both channels of the device in Fig.\ \ref{fig:Device}(b) are in a separable spin state initially. At this stage, there is a high potential barrier between the channels and they are too far apart to interact. We also assume they are in eigenstates of the $z$-axis spin. The spin part of the wave function can thus be labelled $\ket{s_1}\!\ket{s_2}$, meaning that the first electron is in spin state $s_1$, and the second one is in spin state $s_2$.
For a double-dot potential, the two-particle ground state is symmetric in spatial coordinates, described by a spatial wave function $\ket{\Psi^\text{S}(r_1,r_2)}$, while the first excited state is anti-symmetric, with a spatial wave function $\ket{\Psi^\text{A}(r_1,r_2)}$.
We call the spin-antisymmetric combination a singlet state $\ket{\text{S}}$, which corresponds to the ground state with energy $E_\text{S}$, and the symmetric state a triplet state $\ket{\text{T}}$, which is the first excited state with energy $E_\text{T}$ (see Fig.\ \ref{fig:InitialState}):
\begin{equation}
    \ket{\text{S}} = \frac{1}{\sqrt{2}}\ket{\Psi^\text{S}(r_1,r_2)}\left(\ket{\uparrow}\!\ket{\downarrow} - \ket{\downarrow}\!\ket{\uparrow}\right)
\end{equation}
\begin{equation}
    \ket{\text{T}} = \frac{1}{\sqrt{2}}\ket{\Psi^\text{A}(r_1,r_2)}\left(\ket{\uparrow}\!\ket{\downarrow} + \ket{\downarrow}\!\ket{\uparrow}\right)
\end{equation}
We choose the double-dot potential of the gate region such that an equal linear combination of these states has both particles well localized in different channels. This results in the eigenstates of initial high tunnel barrier and those of the gate region having a high overlap. The disturbance introduced by adiabatically changing the tunnel barrier in the SAW reference frame is thus minimized. We can write down combined space and spin states as $\ket{s_1 s_2}_{\text{LR}}$, with particle 1 being in the left channel with spin $s_1$ and particle 2 being in the right channel with spin $s_2$. They are linear combinations of the triplet and singlet states:

\begin{equation}
    \begin{split}
    \ket{\downarrow\uparrow}_{\text{LR}} & = \frac{1}{\sqrt{2}}\left( \ket{\text{T}}+\ket{\text{S}} \right)  \\
    & = \frac{1}{\sqrt{2}}\left(\ket{\Psi^{\text{RL}}(r_1,r_2)}\ket{\uparrow}\!\ket{\downarrow} -
    \ket{\Psi^{\text{LR}}(r_1,r_2)}\ket{\downarrow}\!\ket{\uparrow} \right),
    \end{split}
\end{equation}

\begin{equation}
     \begin{split}
    \ket{\uparrow\downarrow}_{\text{LR}} & =\frac{1}{\sqrt{2}}\left( \ket{\text{T}}-\ket{\text{S}} \right)  \\
    &=\frac{1}{\sqrt{2}}\left(\ket{\Psi^{\text{LR}}(r_1,r_2)}\ket{\uparrow}\!\ket{\downarrow} -
    \ket{\Psi^{\text{RL}}(r_1,r_2)}\ket{\downarrow}\!\ket{\uparrow} \right),
    \end{split}
\end{equation}

where $\ket{\Psi^{\text{LR}}(r_1,r_2)}$ denotes a spatial state with particle 1 in the left channel (negative $x$) and particle 2 in the right channel (positive $x$). These take the form

\begin{equation}
 \ket{\Psi^{\text{RL}}(r_1,r_2)}= \frac{1}{\sqrt{2}}\left(\ket{\Psi^\text{S}(r_1,r_2)} +\ket{\Psi^\text{A}(r_1,r_2)}\right),
\end{equation}
\begin{equation}
 \ket{\Psi^{\text{LR}}(r_1,r_2)}= \frac{1}{\sqrt{2}}\left(\ket{\Psi^\text{S}(r_1,r_2)} -\ket{\Psi^\text{A}(r_1,r_2)}\right).
\end{equation}

A system placed in such a linear superposition oscillates coherently with the period, $~{2\pi\hbar}/{(E_\text{T}-E_\text{S})}$, determined by the energy difference between the ground state and first excited state. A full SWAP operation takes half of this period whilst the root-of-SWAP operation takes a quarter of it, i.e.\ half the duration of a SWAP. In the limit where the on-site Coulomb energy is much greater than the hopping energy, the doubly-occupied states have vanishingly small probability amplitudes and can be ignored \citep{barnes2000quantum}. The state during the time evolution is
\begin{equation}
   \ket{\psi(t)} = \frac{1}{\sqrt{2}}\left\{\ket{\text{T}}+\exp\left(\frac{-it}{\hbar} \Delta E \right)\ket{\text{S}}\right\},
   \label{eq:dynamicsModel}
\end{equation}
where $\Delta E = E_\text{T}-E_\text{S}$.
This description of the power-of-SWAP operation allows us to calculate the probabilities of observing spin-up (spin-down) particles in the left (right) channels after the operation.
The probability of measuring a swapped state, assuming an initial state $\ket{\uparrow \downarrow}_{\text{LR}}$ and a fixed time of interaction $\tau$, is given by
\begin{align}
    \begin{split}
        P_{\text{SWAP}}\left( J \right) = {\mid \braket{\psi(t=\tau)|\downarrow\uparrow}_{\text{LR}}\mid}^2   \\ = \text{sin}^2\left(\frac{1}{2}{J\cdot\tau}\right),
        \label{eq:PoS}
    \end{split}
\end{align}
where $J=\Delta E/(2 \pi \hbar)$.
This probability, given an input state, depends only on the energy difference between the triplet and singlet states, which in turn is a function of the device potential.

\section{Simulations and results}
\label{sec:SimResults}

Numerical simulations of the two-particle dynamics over two dimensions are computationally expensive. However, the ability to model a complete set of energy eigenstates reveals a more complicated behavior, in contrast with analytical two-site models or one-dimensional simulations. In this section, we present the numerical results of entanglement generation via two different root-of-SWAP implementations \citep{burkard1999coupled, barnes2000quantum, owen2012generation}. In both cases, we find that the realistic dynamics deviates from the simpler models.

\subsection{Numerical methods}

The eigenstates of the double-dot system are obtained by numerically solving the Hamiltonian built using the allowed two-particle basis states (Appendix\ \ref{appendix:SecondQuantization}).
To reduce the size of our matrix representation, we can find the initial state of the two-particle system efficiently by using a momentum-space (rather than a position-space) eigensolver (Appendix \ref{appendix:MomentumEigensolver}). Since the system of interest is very close to the ground state, only a small number of momentum basis states are needed. 
The ground and first excited spatial states found using this method are presented in Fig.\ \ref{fig:InitialState}.

These time-independent methods are sufficient to find the initial state of the system and to describe its time evolution in a constant potential. However, when the potential varies as the electron travels across the device, time-dependent simulations need to be used. We evolve the TDSE iteratively using the staggered-leapfrog algorithm \citep{goldberg1967computer, askar1978explicit, maestri2000two}.

Specific details on our GPU implementation of the eigensolver and staggered-leapfrog simulator can be found in Appendix \ref{appendix:GPU}.

\subsection{Coulomb tunneling entanglement generation}

\begin{figure*}[htpb]
\centering
\includegraphics[width=0.95\textwidth]{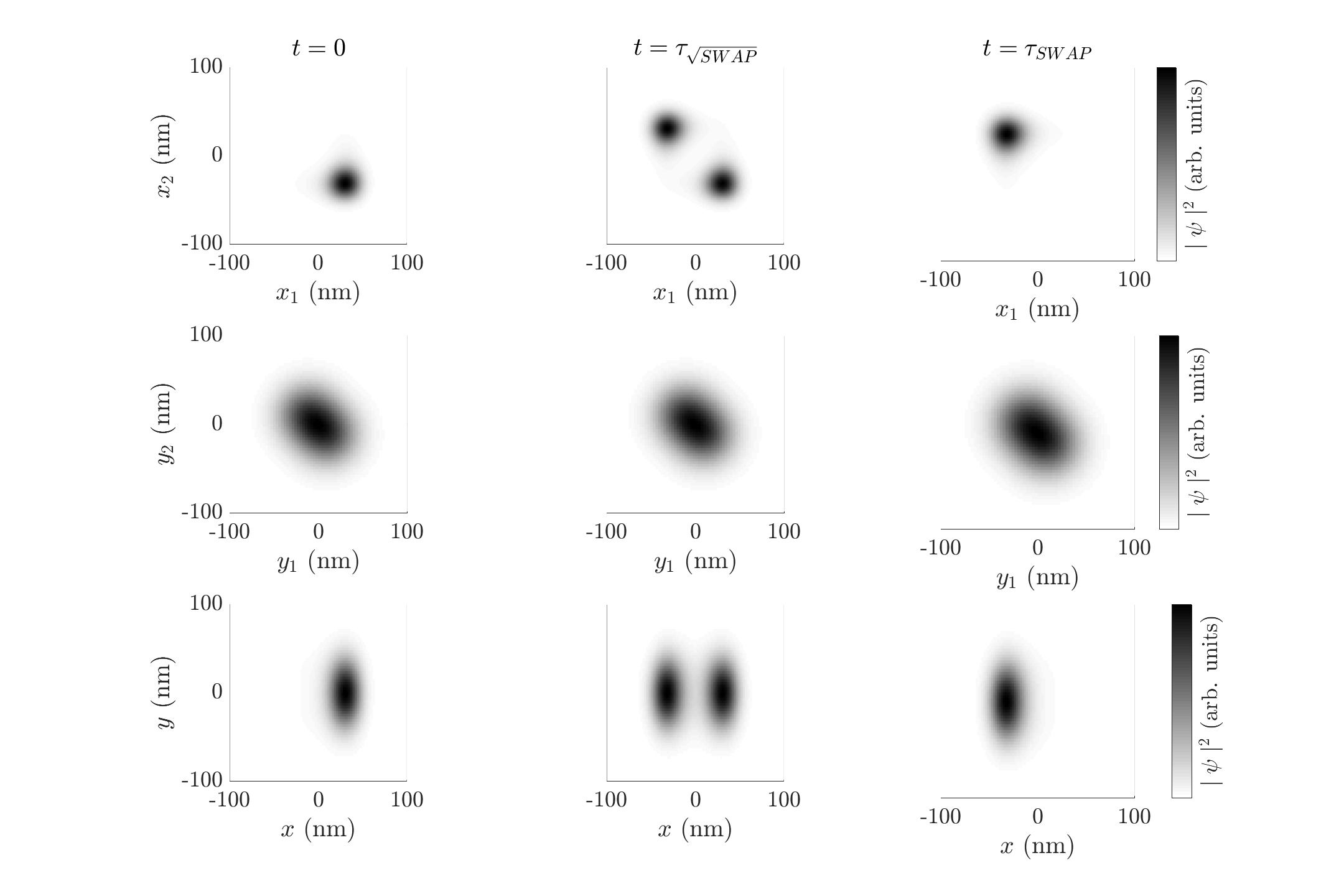}
\caption{\textbf{Entanglement generation using the Coulomb tunneling method.} Top and middle row: trace over the $x$-dimension and $y$-dimension respectively for the initial state (left), root-of-SWAP state (centre), and SWAP state (right) of the wave function.
Bottom row: trace over the second particle for the initial state (left), root-of-SWAP state (centre), and SWAP state (right) of the wave function. Coordinates are chosen to be in the SAW frame of reference with $y = 0$ corresponding to a SAW minimum and $x = 0$ the peak of the tunnel barrier.}
\label{fig:SlowSWAP}
\end{figure*}

Building on a proposal from Ref.\ \citep{barnes2000quantum}, we explore an exchange-interaction based method for the generation of entanglement between two electrons in a SAW system. As described in Section \ref{sec:Analytical}, the two electrons occupy adjacent channels separated by a tunneling barrier, suppressing any wave function overlap. When the electrons enter the low-barrier gate region, they can tunnel through to the other channel at a rate that is determined in part by the barrier height and in part by the Coulomb force, thus allowing for the control of the power-of-SWAP gate by tuning the appropriate Schottky gates. Fig.\ \ref{fig:SlowSWAP} shows snapshots of the wave function during an entangling operation with realistic experimental parameters. When the potential barrier is low, the two-particle state undergoes coherent oscillations between the initial state and the fully swapped state. The duration of the two-particle operation is determined by the length of the tunnel-coupled region. Since the speed of a SAW is constant in a given material, the operation is identical for all incoming electron pairs.

Starting with Eq.\ \ref{eq:PoS}, and assuming that $J$ is exponentially dependent on the tunnel barrier height $A_{\text{TB}}$, and time of interaction $\tau$ is fixed, the probability of the final state being swapped with respect to the initial state has the following dependence on the tunnel barrier:

\begin{align}
    \begin{split}
        P_{\text{SWAP}}\left(A_{\text{TB}}, \tau \right) =  \text{sin}^2\left(\frac{1}{2}{J_0\cdot e^{-b\cdot A_{\text{TB}}}\cdot\tau}\right),
        \label{eq:PoS2}
    \end{split}
\end{align}
where $J_0$ and $b$ are numerically determined parameters. Fig.\ \ref{fig:PowerofSWAP} shows a fit of our time-dependent numerical simulation data (See Appendix \ref{appendix:Parameters} for parameter values used) with the analytical prediction from Eq.\ \ref{eq:PoS2}. It is important to note that although Eq.\ \ref{eq:PoS2} can describe the behaviour of a power-of-SWAP under ideal conditions, a numerical approach is required to account for more realistic scenarios. These can include the presence of impurities in the quantum channels as well as a finite transition length between the low and high tunnel barrier heights.
The inset in Fig.\ \ref{fig:PowerofSWAP} shows the probability amplitude of each computational basis state as the electrons travel through a root-of-SWAP gate. Interactions between the electrons are initially prohibited by the high potential barrier separating them. As they are carried through the interaction region, the electrons become entangled. Upon leaving the region of low potential barrier, the particles can no longer interact and the probability amplitudes become constant.
We find that the SWAP probability around $P_{\text{SWAP}}=0.5$ varies with the tunnel barrier height at a rate of $8.07 \times 10^{-4} ~\mu \text{eV}^{-1}$. This allows for an experimentally viable tunability of the quantum gate via the control of the tunnel barrier height. Assuming a device temperature of 300 mK, tunnel barrier variations due to thermal fluctuations will decrease the root-of-SWAP fidelity by $<0.1\%$. This error could be reduced by increasing the height of the tunnel barrier, at the cost of extending the operation time.

\begin{figure*}[htpb]
\centering
\includegraphics[width=0.85\textwidth]{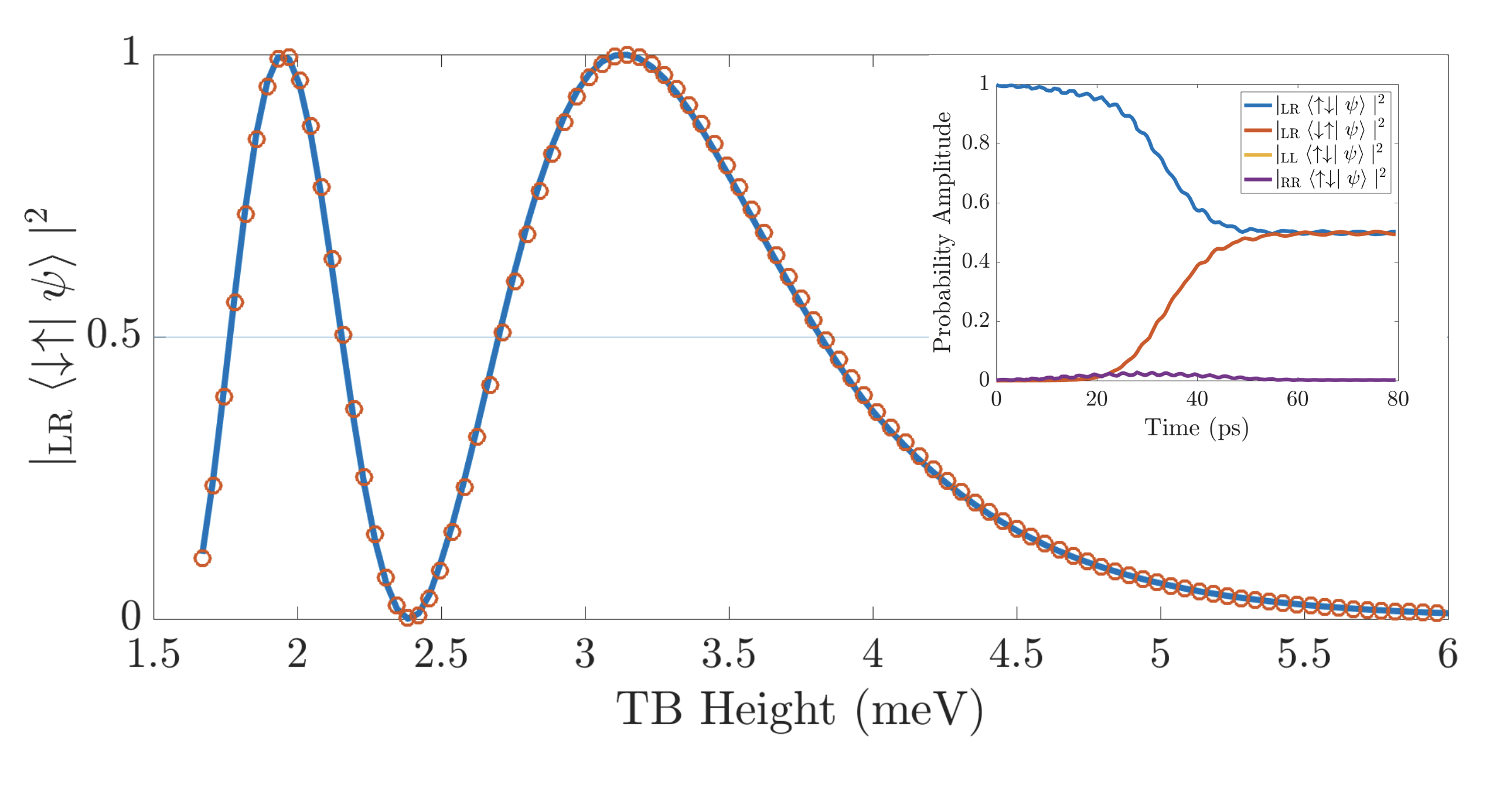}
\caption{\textbf{Probability of SWAP as a function of tunnel barrier height for fixed interaction duration.} Time evolution simulation results (red circles) are fit using Equation \ref{eq:PoS} (solid blue line). The parameters $J_0=2.888 \text{ps}^{-1}$ and $b=0.933 \text{meV}^{-1}$ were found numerically. The inset figure illustrates the occupation of the computational basis states as well as the double occupancy states. In this example, the input state $\ket{\uparrow\downarrow}_{\text{LR}}$ undergoes a root-of-SWAP operation with finite tunnel barrier potential ramps.}
\label{fig:PowerofSWAP}
\end{figure*}

\subsection{Comparison to Models}
To solve the dynamics of the power-of-SWAP operation in our heterostructure SAW-based device, including the 2D spatial extent of the wave function and a time-dependent potential, numerical simulations must be used. However, to avoid lengthy and complicated computations, $\Delta E$ can be estimated using simplified two-site models, thus getting an approximation for the power-of-SWAP extent via Eq.\ \ref{eq:PoS}. 


Assuming a tight-binding-like model, where electrons can tunnel between the quantum dots, we can estimate the full 2D time evolution by applying the Hund-Mulliken model for molecular orbitals \citep{burkard1999coupled} (see Appendix\ \ref{appendix:HMModel}).
Alternatively, the evolution of the two-particle state can also be modelled with the Hubbard approach for short range Coulomb interaction \cite{barnes2000quantum} (see Appendix\ \ref{appendix:HubbardModel}).
Solving the Hund-Mulliken Hamiltonian, we find the eigenenergies associated with the singlet and triplet states and define the SWAP frequency in terms of the on-site energy $U$ and the hopping term $t_{\text{h}}$:

\begin{equation}
\label{eq:HMInText}
    J=\frac{1}{2\pi\hbar}\left[V_{-} - V_{+} +\frac{1}{2}\left( \sqrt{U_{\text{h}}^2+16t_{\text{h}}^2}-U_{\text{h}}\right)\right]=\frac{\Delta E}{2\pi\hbar},
\end{equation}
where $U_{\text{h}}=U - V_{+} + X$.

For the simplified Hubbard Hamiltonian, this expression reduces to
\begin{equation}
\label{eq:HubbardInText}
    J=\frac{1}{4\pi\hbar}\left(-U+\sqrt{U^2+16t_{\text{LR}}^2}\right)=\frac{\Delta E}{2\pi\hbar}.
\end{equation}

For realistic Hamiltonians, it is impossible to obtain $U$ analytically. Instead, we numerically calculate this parameter. To avoid unphysical results introduced by the $1/r$ factor, a softened Coulomb potential is used \citep{owen2012generation} both in the models and the numerical simulations throughout this work. We implement this softening by assuming that the wave function has a Gaussian spread in the third dimension, with a standard deviation of $\Delta_z$.

\begin{figure}[htp]
\centering
\includegraphics[width=0.5\textwidth]{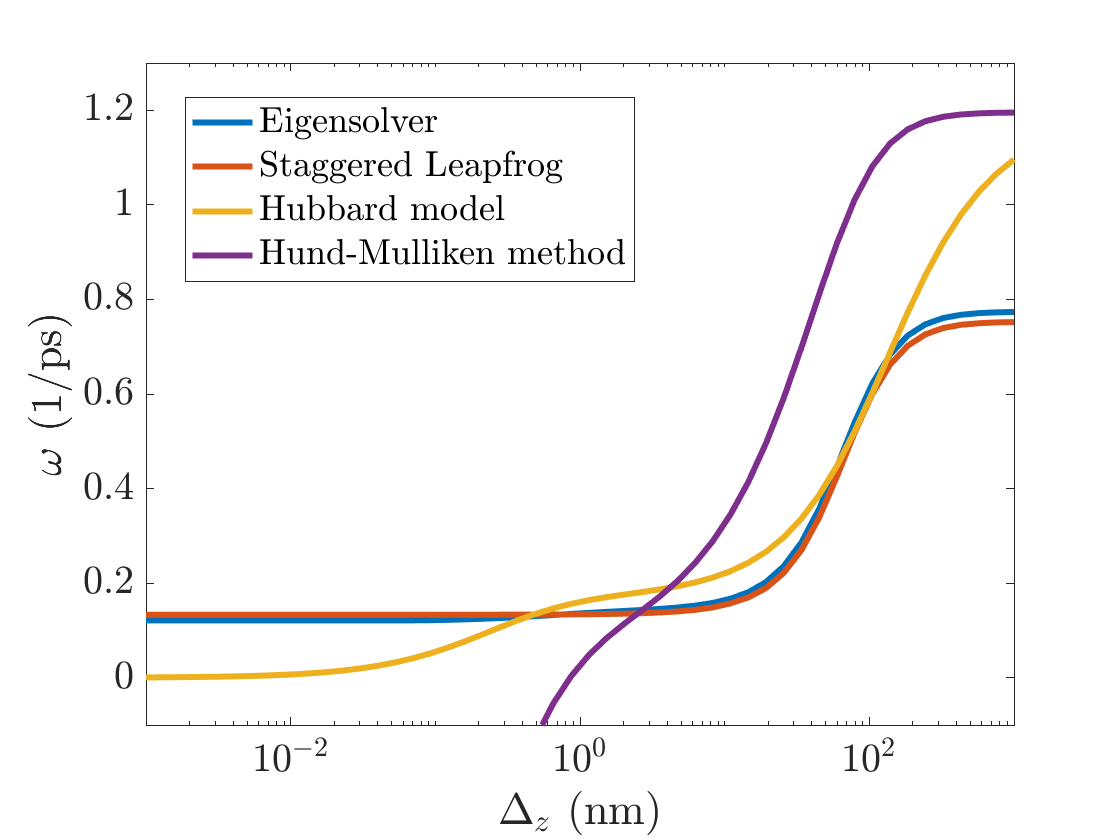}
\caption{\textbf{Comparison to analytical models.} Power-of-SWAP frequency as a function of effective wave function spread in the $z$-dimension. Coulomb softening accounts for the finite $z$-dimension and plays an important role in determining the rate of the exchange interaction.}
\label{fig:4models}
\end{figure}

We compare both the Hubbard model and the Hund-Mulliken method described above to our simulation results for a range of $\Delta_z$. We find a close match between the frequency of the SWAP operation as calculated by our time-dependent numerical solver and these obtained by solving the eigenvalue problem directly. Both models (Eq. \ref{eq:HMInText} and Eq.\ \ref{eq:HubbardInText}) show significant discrepancy for most values of $\Delta_z$. Moreover, the Hund-Mulliken model predicts negative frequencies for $\Delta_z < 1$  nm. We conclude that although both models provide a reasonable qualitative prediction of the two-particle dynamics for Gaussian spread of $\Delta_z \sim10-100$ nm, a more sophisticated numerical approach, such as the one used in this work, is required to obtain precise quantitative dynamics. A comparison of both analytical models with our simulations can be seen in Fig.\ \ref{fig:4models}.

\subsection{Entanglement generation via electron collisions}
\label{sec:Collision}

In a previously suggested root-of-SWAP scheme \citep{owen2012generation}, two electrons are travelling in individual channels separated by a high potential barrier, such that there is no wave-function overlap. The potential barrier \textit{abruptly} (or diabatically) changes in the SAW reference frame such that the two channels are joined to create a global potential minimum between them. Without the presence of the barrier, both electrons fall towards one another in a harmonic oscillator potential and interact via the Coulomb force. Once the operation is completed, the central barrier is reintroduced, causing the reappearance of separate decoupled channels. As the quantum states of particles in layered semiconductor technologies are confined in the dimension perpendicular to the quantum wells, which has a constant potential throughout the device, the third dimension does not significantly affect the operation. However, we find that the previous reduction to 1D is an oversimplification, as the possible spatial dynamics in the second dimension strongly affects the electron-electron interactions.

Here, we simulate this $\textit{single-shot}$ (i.e. in a single collision) entanglement generation, and we find that under current experimentally realistic parameters, it is impossible to generate a root-of-SWAP, or any significant entanglement over the $x$-dimension.

\begin{figure*}[htpb]
\centering
\includegraphics[width=0.95\textwidth]{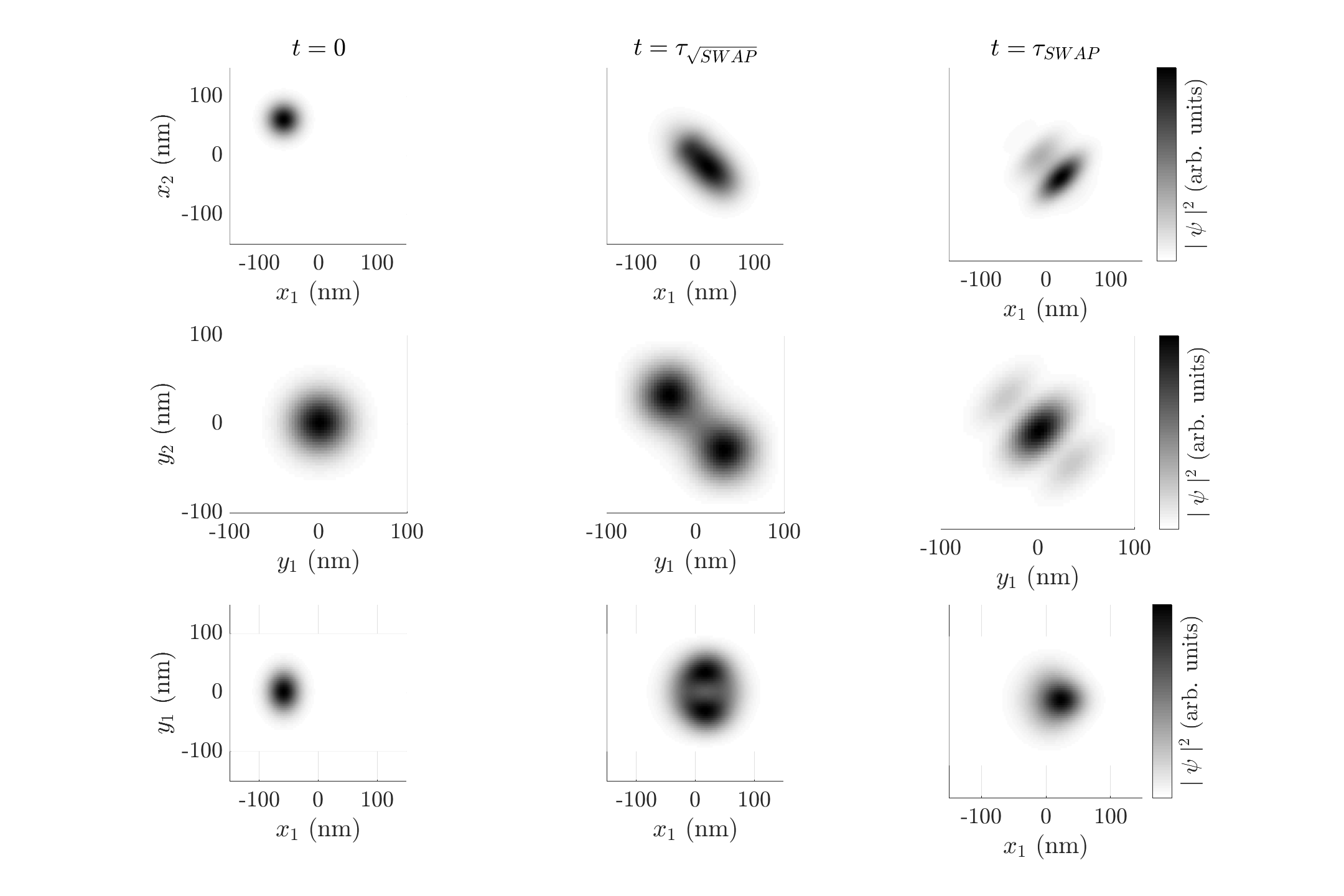}
\caption{\textbf{Entanglement generation via the collision of two electrons.} Top and middle row: trace over the $x$-dimension and $y$-dimension respectively for the initial state (left), root-of-SWAP state (centre), and SWAP state (right) of the wave function.
Bottom row: trace over the second particle for the initial state (left), root-of-SWAP state (centre), and SWAP state (right) of the wave function.
 Coordinates are chosen to be in the SAW frame of reference with $y = 0$ corresponding to a SAW minimum and $x = 0$ the middle of the harmonic channel.}
\label{fig:CollisionMethod}
\end{figure*}

Fig.\ \ref{fig:CollisionMethod} shows snapshots of the two-electron wave function undergoing a single collision in two dimensions. The wave function remains fully separable along the $x$-dimension. However, in the $y$-dimension, it transitions from a Gaussian-like low-energy state to a more spread-out entangled state. This is conflicting with the desired outcome of generating a maximally entangled state in the $x$-dimension. The operation is effectively a SWAP instead of a root-of-SWAP, with the additional downside of exciting higher-energy states in the $y$-dimension.
These unwanted spatial excitations of the wave function lead to lowering the spatial fidelity of the operation and thus it is not possible to concatenate multiple operations for useful quantum information processing. This also prevents the restoration of the wave function to its original state by applying the SWAP twice, a fundamental property of this operation.
We find that increasing the $y$ confinement does not prevent this behaviour until the SAW amplitude is increased by a factor of order $10^3$, where the problem effectively reduces to 1D. However, this would require SAW amplitudes on the order of $10^4$ meV, which is experimentally unrealistic \citep{talyanskii1997single}. Varying the $x$ confinement over a wide range also does not solve the issue. Therefore we conclude that the collision method is unable to produce the root-of-SWAP operation in a realistic 2D scenario.

\section{Experimental Sensitivity}
\label{sec:Sensitivity}

 Here, we investigate the power-of-SWAP operation's sensitivity to disturbances in  $A_{\text{TB}}$ and $\tau$, which the output probabilities depend on. From an information-theoretical perspective, an experiment's sensitivity to an unknown physical parameter, $\theta$, is quantified by the Fisher information:
\begin{equation}
    \label{eq:ClassFisher}
F (\theta)=\sum_{i } P(\mathcal{M}_i | \theta)  \bigg[ \frac{\partial}{\partial \theta} \ln P(\mathcal{M}_i | \theta) \bigg]^2  ,
\end{equation}
where $\mathcal{M}_i$ denotes the $i^{\mathrm{th}}$ measurement's outcome \citep{helstrom1969quantum}. Given $N$ experimental runs, the precision of an estimate $\hat{\theta}$ of $\theta$ is bounded by the Cram\'er-Rao inequality: $\mathrm{Var}( \hat{\theta} ) \geq [N F(\theta)]^{-1}$, such that the greater the Fisher information, the smaller the estimator's variance can be \citep{frieden2004science}. Using the output probabilities from Sec. \ref{sec:Analytical}, we find that $F(A_{\text{TB}}) =  b^2 \tau^2 J_0^2 e^{-2bA_{\text{TB}}}$: the ability to estimate $A_{\text{TB}}$ decreases exponentially with $A_{\text{TB}}$ itself. However, for parameters that yield the root-of-SWAP operation ($A_{\text{TB}} \approx 3.86 \; \rm{meV}$) we find that $F(A_{\text{TB}}) \approx 2.15 \; (\rm{meV})^{-2}$. This value of $F(A_{\text{TB}})$ lower bounds the standard deviation of $A_{\text{TB}}$: $\sigma_{A_{\text{TB}}} \gtrapprox 0.012 \; \rm{meV}$ in an experiment with $N=3000$ trials. Despite the exponentially decreasing sensitivity of tunnel-barrier heights, the relevant values for a root-of-SWAP operation are within the experimentally viable regime \citep{Takada2019} specified in Appendix B. The Fisher information about $\tau$ is $ F( \tau ) = J_0^2 e^{-2bA_{\text{TB}}} $ is constant with respect to $\tau$ itself. For the ideal root-of-SWAP parameters, we find that $F(\tau) \approx 6.17 \times 10^{-3} \; \rm{ps}^{-2}$, which gives a lower bound on the standard deviation of $\tau$: $\sigma_{\tau} \gtrapprox 0.23 \; \rm{ps}$, when $N=3000$.

\section{Discussion And Conclusion}
\label{sec:Conclusion}

The two-qubit entangling operation is an essential building block of a quantum information processor. We have shown that surface-gate-controlled flying electron-spin qubits are able to generate entanglement through the power-of-SWAP operation in a reliable and stable fashion. We show that the problem of wave function dispersion can be solved through the use of SAWs, which generate the potential confinement needed to preserve the wave function's profile. We present accurate numerical solutions to the time-dependent Schr\"{o}dinger equation using a staggered-leapfrog method and we investigate previously proposed schemes for generating entanglement between electron-spin qubits.
 We find that realising the power-of-SWAP operation via electron collision \citep{owen2012generation} suffers from significant problems, whilst an implementation based on tunnelling \citep{barnes2000quantum,burkard1999coupled} is shown to be realizable with high fidelity even when experimental control of the tunneling barrier is imperfect.
We find that this entangling operation governed by the exchange interaction and coherent tunneling of electrons offers a more stable approach and makes high gate fidelities possible.

Our two-particle simulations use experimentally realistic parameters and potential layouts and show that such devices are readily realizable using current semiconductor fabrication techniques.
While the behaviour of an ideal system can be predicted exactly by solving the Hamiltonian and assuming that the electrons are initialized to and remain in a combination of triplet and singlet states, the advantage of our numerical methods is to simulate realistic entangling operations.
Although these simulations were focused on the experimental parameters of GaAs-based devices, the same behaviour is expected in other SAW-based semiconductor devices. Moreover, our findings can be generalized to systems that do not include SAWs. Static quantum dots, confined in every dimension and separated by a tunnel barrier interact in the same way. Such a tunnel barrier can be modulated using fast microwave pulses \citep{Brunner2011} to reproduce the two-particle dynamics presented in this paper. A static root-of-SWAP gate was recently realised with high fidelity using phosphorus donors in silicon \citep{He2019}, proving that such systems are achievable experimentally. Coherent spin state SWAP operations between electron-spin qubits in a quadruple array of semiconductor quantum dots were also lately achieved \citep{Kandel2019}.

Our results provide new evidence that an entangling root-of-SWAP gate based on the exchange interaction is experimentally viable in SAW-based semiconductor heterostructures.

\section{Acknowledgements}
A.A.L. and H.V.L. have contributed equally to this work. This project has received funding from the European Union's Horizon 2020 research and innovation programme under the Marie Sk\l{}odowska-Curie grant agreement No 642688. This work was supported by the Project for Developing Innovation Systems of the Ministry of Education, Culture, Sports, Science and Technology (MEXT), Japan. A.A.L. acknowledges support from EPSRC and Hitachi via CASE studentships RG 94632. D.R.M.A.S. acknowledges support from the EPSRC, Hitachi Cambridge, Lars Hierta's Memorial Foundation and the Sweden-America Foundation. The authors would also like to thank Prof.\ Chris Ford for valuable discussions and comments.

\clearpage
\appendix

\section{Power-of-SWAP quantum logic gate}
\label{appendix:PowerOfSwap}

In the two-qubit basis $\ket{00} , \ket{01}, \ket{10}, \ket{11}$, the Power-of-SWAP operation for $n-th$ power is represented by the matrix:

\begin{equation}
SWAP^n =
    \begin{pmatrix}
    1 & 0 & 0 & 0 \\
    0 & \frac{1}{2}(1 + e^{i \pi n}) & \frac{1}{2}(1 - e^{i \pi n})  & 0 \\
    0 & \frac{1}{2}(1 - e^{i \pi n}) & \frac{1}{2}(1 + e^{i \pi n}) & 0 \\
    0 & 0 & 0 & 1 \\
    \end{pmatrix},
    \label{eq:HamiltonianSwapN}
\end{equation}

For root-of-SWAP, $n=\frac{1}{2}$ and the matrix representation is:

\begin{equation}
\sqrt{SWAP} =
    \begin{pmatrix}
    1 & 0 & 0 & 0 \\
    0 & \frac{1}{2}(1 + i) & \frac{1}{2}(1 - i)  & 0 \\
    0 & \frac{1}{2}(1 - i) & \frac{1}{2}(1 + i) & 0 \\
    0 & 0 & 0 & 1 \\
    \end{pmatrix},
    \label{eq:HamiltonianSwapRoot}
\end{equation}

\section{Parameter values}
\label{appendix:Parameters}

\begin{center}
\begin{table}[h!]
 \begin{tabular}{l l}
 \hline\hline
 Parameter & Value range \\ [0.5ex]
 \hline
 Distance between channels  & 80 nm   \\

 Tunnel coupled region start  & $y_d$ = 36 nm   \\
 
 Tunnel coupled region end & $y_u$ = 144 nm   \\

 Interaction time & $\tau = $ 36 ps   \\

 SAW amplitude & $A_{\text{SAW}}$ = 25 meV  \\

  SAW wavelength & $\lambda = 1 \upmu$m  \\

  SAW velocity & $v$ = 3 nm / ps  \\

  Harmonic channel confinement & $\omega_x^2$ = 0.002 $\frac{\text{meV}}{{\text{nm}}^2 m_e}$ \\
  
 Electron effective mass & 0.067 $m_e$ \\

 Relative permitivitty (GaAs) & 13.1  \\

 Gaussian tunneling barrier amplitude & $A_1$ = 15.3  meV \\

 Gaussian tunneling barrier width  & $\sigma_1$ = 30 - 40 nm \\ 
 
 Gaussian barrier amplitude & $A_2$ = 510  meV \\
 
 Gaussian barrier width & $\sigma_2$ = 0.8  nm \\
 
 Transition between barrier heights  &  $\sigma_y$ = 10  nm \\
 
 Coulomb softening  &  $\Delta_z$ = 10 - 100  nm \\[1ex]
 \hline\hline
\end{tabular}
\caption{Ranges of parameter values used in simulations.}
\label{table:1}
\end{table}
\end{center}

Explicit form of potentials used in Eq. \ref{eq:HamiltonianDevice}, in terms of the parameters above, in reference frame of the SAW:
\begin{align}
    \begin{split}
    &V_{\text{D}}(x,y) = \frac{m_e}{2} \omega_x^2 x^2 
    + A_{1} \exp \left(\frac{-x^2}{2\sigma_{1}^2} \right) + \frac{A_{2}}{2} \exp \left(\frac{-x^2}{2\sigma_{2}^2} \right) \\
    & 
     \times \left(2-\tanh\left(\frac{y-y_d}{\sigma_y}\right)-\tanh\left(-\frac{y-y_u}{\sigma_y} \right) \right).
     \end{split}
\end{align}

\begin{equation}
    V_{\text{SAW}}(x, t) = \frac{A_{\text{SAW}}}{2}\left(1- \cos\left(\frac{x-t v}{\lambda} \right) \right)
\end{equation}

\begin{equation}
    V_{C}(r) =  \frac{e^2}{4  \sqrt{2} \pi \epsilon  \Delta_z U(-\frac{1}{2},0,\frac{r^2}{2 \Delta_z^2})}
\end{equation}

where $U$ is the confluent hypergeometric function of the second kind, which encapsulates a Gaussian spread with standard deviation $\Delta_z$ in the $z$-dimension:

\begin{equation}
    U(a,b,z)=\frac{1}{\Gamma(a)} \int_0^\infty e^{-zt} t^{a-1} (1+t)^{b-a-1} dt
\end{equation}

\section{Second Quantization basis}
\label{appendix:SecondQuantization}

The full set of basis states of the system using second quantization, with two spin-$\sfrac{1}{2}$ fermions occupying  $i$-th and $j$-th out of $N$ spatial sites respectively, is given by
\begin{align}
    \begin{split}
        \ket{\uparrow\uparrow}_{ij} = c^{\dagger}_{i\uparrow}c^{\dagger}_{j\uparrow}\ket{0}, i\neq j\\
        \ket{\downarrow\downarrow}_{ij} = c^{\dagger}_{i\downarrow}c^{\dagger}_{j\downarrow}\ket{0}, i\neq j \\
        \ket{\uparrow\downarrow}_{ij} = c^{\dagger}_{i\uparrow}c^{\dagger}_{j\downarrow}\ket{0}, \\
        \ket{\downarrow\uparrow}_{ij} = c^{\dagger}_{i\downarrow}c^{\dagger}_{j\uparrow}\ket{0}.
    \end{split}
    \label{eq:basis}
\end{align}

These fermionic creation operators obey the anticommutation relation $\{c^{\dagger}_{i s_1}, c^{\dagger}_{j s_2} \} = 0$. Therefore, the basis states are also related by $\ket{\downarrow\uparrow}_{ij} = -\ket{\uparrow\downarrow}_{ji}$.
With $N=2$, the basis states of the Hubbard two-site model as in Appendix \ref{appendix:HubbardModel} are obtained.

\section{Momentum Space Eignesolver}
\label{appendix:MomentumEigensolver}

A real-space wave function can be written in the discrete case as a sum of the momentum eigenfunctions:
\begin{equation}
\ket{\psi(\boldsymbol{r})}=\sum_{k}^{N_k} \phi_k \ket{\psi_{k}(\boldsymbol{r})}
\end{equation}
where $\ket{\psi_{k}(\boldsymbol{r})}$ are the momentum eigenfunctions.\\
Owing to the canonical relation between momentum and position operators,

\begin{equation}
\hat{p}=-i\hbar\frac{\partial}{\partial r},
\label{canonical}
\end{equation}
 the eigenfunctions are

\begin{equation}
\ket{\psi_{k}\left(\boldsymbol{r}\right)}={(2\pi)^{\frac{d}{2}}} \ket{e^{i\boldsymbol{k}\cdot \boldsymbol{r}}}
\end{equation}
in $d$ dimensions. Therefore, the real space wave function can be written as:

\begin{equation}
\ket{\psi(\boldsymbol{r})}={(2\pi)^{\frac{d}{2}}} \sum_{k}^{N_k} \phi_k  \ket{e^{i\boldsymbol{k}\cdot \boldsymbol{r}}  }
\end{equation}

This is just a Discrete Fourier Transform:

\begin{equation}
\ket{\psi(\boldsymbol{r})} = \mathcal{F}\left( \ket{\phi(\boldsymbol{k})}\right)
\end{equation}

Similarly, the inverse is also true:
\begin{equation}
\ket{\phi(\boldsymbol{k})}= \mathcal{F}^{-1}\left(\ket{\psi(\boldsymbol{r})}  \right)
\end{equation}
where $\mathcal{F}^{-1}$ denotes the inverse Discrete Fourier transform. \\

Momentum space can be discretised in multiples of $\frac{2\pi}{L}$, where $L$ is a real space extent over which the amplitude of the wave function should have decayed to 0 near the edges. Defining some cutoff $N_k$, momentum will take the values:

\begin{equation}
{k}_n= n \frac{2\pi}{L},
\end{equation}
where $n=[-N_k,-N_k+1,...,0,1,...,N_k]$.

When we Fourier transform the Schr{\"o}dinger equation  into momentum space, the kinetic term $-\frac{\hbar^2}{2m} \nabla^2$ will become $\frac{\hbar^2 }{mL^2}\left(\text{cosh}(kL) -1 \right)$ owing to discreteness and the canonical momentum-position relation (Eq. \ref{canonical}). The Fourier-transformed potential matrix element becomes

\begin{align}
    \begin{split}
\hat{V}_{kq}={(2\pi)}^{\frac{d}{2}} \sum_{r}^N  e^{-i\boldsymbol{q}\cdot \boldsymbol{r}}  V(\boldsymbol{r})   e^{i\boldsymbol{k}\cdot \boldsymbol{r}}= \\
  {(2\pi)}^{\frac{d}{2}}  \sum_{r}^N   V(\boldsymbol{r})   e^{i(\boldsymbol{k-q})\cdot \boldsymbol{r}} = \mathcal{F} \left(V(\boldsymbol{r}) \right)(k-q)
    \end{split}
\end{align}

The elements of the Hamiltonian matrix  to be solved by diagonalisation in this method take the form
\begin{equation}
H_{kq} = \frac{\hbar^2 }{mL^2}\left[\text{cosh}(kL) -1 \right]\delta_{kq} + \hat{V}_{kq}
\end{equation}
Note that this matrix is not sparse as was the case in real space. This could be thought of as reducing the memory space (smaller matrix) at the cost of computational complexity.  However owing to the efficiency of Fast Fourier Transform algorithms, the problem is faster to solve as opposed to a real-space solver. This method extends readily to two or more particles.

\section{GPU Implementation}
\label{appendix:GPU}

\subsection{Eigensolver}

For the double-quantum-dot system investigated here, time-independent solutions converge for 10 or more momentum basis states in each dimension. 
The momentum wave function is zero-padded to the required real-space number of points, which should be at least 100 in each dimension for the real-space time-dependent solver.
An inverse Discrete Fourier Transform (DFT) is applied. Using the momentum basis makes the two-particle problem tractable in two dimensions, as the Hamiltonian is reduced from $\sim10^{16}$ to $\sim10^8$ elements. Such a matrix of complex floating point numbers reaches the limitations of random access memory (RAM) normally available to modern computers. Our eigensolver can also run on a GPU for increased speed, however GPU Video RAM tends to be smaller than CPU RAM. Memory is a limiting factor, and only state-of-the-art devices are able to solve the problem accurately, i.e. using at least 10 momentum modes in each dimension.
A two-particle 2D problem with 10 points in each dimension can be solved in tens of minutes on a modern desktop computer and gives a result accurate enough to be used as a starting point for the time-dependent solver while keeping the normalization errors below 1\%.

\subsection{Staggered-lepfrog}

We run our time-dependent iterative solver on GPU hardware to accelerate the computation by parallelizing updates for each spatial site. The wave function can be evolved either in real  or momentum space, using DFTs to transform between the two. We find that if the wave function does not contain significant contributions from high-energy eigenstates, employing the momentum basis allows us to reduce the wave-function size compared to real space, whilst keeping the same accuracy. We use about 100 real-space, or 20 momentum-space points, per dimension. However, the momentum solver tends to perform poorly when sudden and strong interactions take place. We find it optimal to use real space for simulating the collision method, and momentum space for the Coulomb tunneling method.

For improved simulation speed and accuracy, we time-evolve only the  $ \ket{\uparrow\downarrow}_{ij}$ part as a spin-independent wave function $\ket{\psi}_{ij}$.
Using such a state is justified, as the root-of-SWAP operation only has an effect on $\ket{\uparrow\downarrow}_{ij}$ and $\ket{\downarrow\uparrow}_{ij}$ states, which are related by a spatial index exchange $(\ket{\downarrow\uparrow}_{ij} = -\ket{\uparrow\downarrow}_{ji})$. Furthermore, $\ket{\uparrow\uparrow}_{ij}$ and $\ket{\downarrow\downarrow}_{ij}$ states are invariant under the power-of-SWAP operation, and thus irrelevant in this work.

\section{Hund-Mulliken model for molecular orbitals}
\label{appendix:HMModel}

The Hund-Mulliken model for molecular orbitals \citep{burkard1999coupled} builds a two-particle basis from right- and left-localized single-particle states $\ket{\phi_{\pm}}$. These states are orthonormalised to $\ket{\Phi_{\pm}} = (\ket{\phi_{\pm}} -g \ket{\phi_{\mp}})/(\sqrt{1-2Sg+g^2})$, where $S=\braket{\phi_{\pm}\mid{\phi_{\mp}}}$ is the wave function overlap and $g=(1-\sqrt{1-S^2})/S$. The singly- and doubly-occupied two-particle basis is constructed with direct products:
\begin{equation}
    \begin{split}
    &\ket{\Psi^{s}_{\mp}} = \frac{1}{\sqrt{2}}\left(\ket{\Phi_{+}} \!\ket{\Phi_{-}} \mp  \ket{\Phi_{-}}\! \ket{\Phi_{+}}\right),   \\
    &\ket{\Psi^{d}_{\mp}} = \ket{\Phi_{\mp}}\! \ket{\Phi_{\mp}}.
    \end{split}
\end{equation}

The Hamiltonian in this basis has the form:
\begin{equation}
\hat{H} =
    \begin{pmatrix}
    V_{-} & 0 & -\sqrt{2}t_{\text{h}} & 0 \\
    0 & V_{+} & -\sqrt{2}t_{\text{h}} & 0 \\
    0 & -\sqrt{2}t_{\text{h}} & U & X \\
    0 & -\sqrt{2}t_{\text{h}} & X & U \\
    \end{pmatrix},
    \label{eq:HamiltonianHM}
\end{equation}
where each entry is defined as:
\begin{equation}
    U = \frac{e^2}{4\pi\epsilon} \bra{\Psi^{d}_{\pm}}\frac{1}{r}\ket{\Psi^{d}_{\pm}},
    \label{eq:OnSiteHund}
\end{equation}
\begin{equation}
    X = \frac{e^2}{4\pi\epsilon} \bra{\Psi^{d}_{\pm}}\frac{1}{r}\ket{\Psi^{d}_{\mp}},
\end{equation}
\begin{equation}
    V_{+} = \frac{e^2}{4\pi\epsilon} \bra{\Psi^{s}_{+}}\frac{1}{r}\ket{\Psi^{s}_{+}},
\end{equation}
\begin{equation}
    V_{-} = \frac{e^2}{4\pi\epsilon} \bra{\Psi^{s}_{-}}\frac{1}{r}\ket{\Psi^{s}_{-}},
\end{equation}

and $t_{\text{h}}$ is the hopping term
\begin{equation}
    t_{\text{h}} = \bra{\Phi_{\pm}}\frac{\hat{p}^2}{2m}\ket{\Phi_{\mp}}- \frac{e^2}{4\sqrt{2}\pi\epsilon} \bra{\Psi^{s}_{+}}\frac{1}{r}\ket{\Psi^{d}_{\pm}}.
\end{equation}

Solving the Hund-Mulliken Hamiltonian in Eq.\ \ref{eq:HamiltonianHM}, we find the eigenenergies associated with the singlet and triplet states and define the SWAP frequency in terms of $U$ and $t_{\text{h}}$:

\begin{equation}
    J=\frac{1}{2\pi\hbar}\left[V_{-} - V_{+} +\frac{1}{2}\left( \sqrt{U_{\text{h}}^2+16t_{\text{h}}^2}-U_{\text{h}}\right)\right]=\frac{E_\text{T}-E_\text{S}}{2\pi\hbar},
\end{equation}
where $U_{\text{h}}=U - V_{+} + X$.

\section{Hubbard model}
\label{appendix:HubbardModel}

The evolution of the two-particle state can be modelled with the Hubbard approach for short range Coulomb interaction without magnetic fields \cite{barnes2000quantum}. The simplified 2-site Hamiltonian in the second quantization basis from Eq.\ \ref{eq:basis} then has the form
\begin{equation}
\hat{H} =
    \begin{pmatrix}
    V & 0 & 0 & 0 & 0 & 0 \\
    0 & V & 0 & 0 & 0 & 0 \\
    0 & 0 & 0 & 0 & -t_{\text{LR}} & -t_{\text{LR}} \\
    0 & 0 & 0 & 0 & t_{\text{LR}} & t_{\text{LR}} \\
    0 & 0 & -t_{\text{LR}} & t_{\text{LR}} & U & 0 \\
    0 & 0 & -t_{\text{LR}} & t_{\text{LR}} & 0 & U \\
    \end{pmatrix},
    \label{eq:HamiltonianBarnes}
\end{equation}

where $t_{\text{LR}}$ is the hopping term
\begin{equation}
    t_{\text{LR}} =  \tensor[_{\text{L}}]{\bra{\uparrow}}{}\frac{\hat{p}^2}{2m}\ket{\uparrow}_{\text{R}} = \tensor[_{\text{L}}]{\bra{\downarrow}}{}\frac{\hat{p}^2}{2m}\ket{\downarrow}_{\text{R}},
\end{equation}
$U$ is the on-site energy
\begin{equation}
    U = \frac{e^2}{4\pi\epsilon } \tensor[_{\text{{LL}}}]{\bra{\uparrow\downarrow}}{}\frac{1}{r}\ket{\uparrow\downarrow}_{\text{{LL}}} = \tensor[_{\text{{RR}}}]{\bra{\uparrow\downarrow}}{}\frac{1}{r}\ket{\uparrow\downarrow}_{\text{{RR}}},
    \label{eq:OnSite}
\end{equation}
and
\begin{equation}
    V = \frac{e^2}{4\pi\epsilon} \tensor[_{{\text{LR}}}]{\bra{\uparrow\uparrow}}{}\frac{1}{r}\ket{\uparrow\uparrow}_{\text{{LR}}} = \tensor[_{\text{{LR}}}]{\bra{\downarrow\downarrow}}{}\frac{1}{r}\ket{\downarrow\downarrow}_{\text{{LR}}}.
\end{equation}

Solving this Hamiltonian, we find the eigenenergies associated with the singlet and triplet states and define the SWAP frequency in terms of $U$ and $t_{\text{LR}}$:

\begin{equation}
    J=\frac{1}{4\pi\hbar}\left(-U+\sqrt{U^2+16t_{\text{LR}}^2}\right)=\frac{E_\text{T}-E_\text{S}}{2\pi\hbar}.
\end{equation}

\clearpage

\providecommand{\noopsort}[1]{}\providecommand{\singleletter}[1]{#1}%

\end{document}